\documentclass[fleqn,twoside]{article}
\usepackage[headings]{espcrc2}

\usepackage{graphicx}
\usepackage[figuresright]{rotating}
\usepackage{pstricks}
\usepackage{amsmath}

\newcommand{\be}{\begin{equation}}
\newcommand{\ee}{\end{equation}}
\newcommand{\bea}{\begin{eqnarray}}
\newcommand{\eea}{\end{eqnarray}}
\newcommand{\setl}{\setlength\arraycolsep{2pt}}

\newcommand{\ra}{\rightarrow}

\newcommand{\lesssim}{ {\
\lower-1.2pt\vbox{\hbox{\rlap{$<$}\lower5pt\vbox{\hbox{$\sim$}}}}\ } 
}
\newcommand{\gtrsim}{ {\
\lower-1.2pt\vbox{\hbox{\rlap{$>$}\lower5pt\vbox{\hbox{$\sim$}}}}\ } 
}

\newcommand{\cO}{{\cal O}}

\newcommand{\Ree}{\mbox{\rm Re}}

\newcommand{\MeV}{\mbox{\rm MeV}}
\newcommand{\GeV}{\mbox{\rm GeV}}

\newcommand{\with}{\mbox{\rm with}}

\newcommand{\gL}{\frac{1-\gamma_{5}}{2}}
\newcommand{\gR}{\frac{1+\gamma_{5}}{2}}

\title{The L-R Correlator and its Chiral Condensates in the MHA and MHA+V' Approximations to Large--$N_c$ QCD}

\author{ Samuel Friot \address[CPT]{Centre  de Physique Th{\'e}orique~CNRS-Luminy, Case 907 
F-13288 Marseille Cedex 9, France \\ CNRS UMR 6207} \address[LM]{Laboratoire de Math\'ematiques Universit\'e 
Paris-Sud - B\^at. 425, 91405 Orsay Cedex, France \\ CNRS UMR 8628}\thanks{Talk given at the $11^{th}$ 
High-Energy Physics International Conference on Quantum Chromodynamics, 5-10 July (2004), Montpellier 
(France).}}

\runtitle{The L-R Correlator and its Chiral Condensates in the MHA and MHA+V' Approximations...}
\runauthor{S. Friot}

\begin{document}

\begin{abstract}
In this talk, we describe part of a recent work \cite{FGdeR04} on the correlation function of a  $V-A$ current with a  
$V+A$ current in the framework of QCD in the limit of a large number of colours $N_c$. The discussion takes place 
within two successive approximations of this theory, called MHA and MHA+V'. Results concerning the evaluation of 
chiral condensates of dimension six and eight, as well as matrix elements of the $Q_7$ and $Q_8$ electroweak 
penguin operators, are given. 
\vspace{1pc}
\end{abstract}
\maketitle
\section{Introduction}\label{int}

The correlation function of a left--handed current
 with a right--handed current
\vspace{-0.2cm}
\begin{equation}
\nonumber
L^{\mu}(x)=\bar{u}(x)\gamma^{\mu}\gL d(x) 
\end{equation}
and
\begin{equation}
\nonumber
R^{\nu}(0)=\bar{d}(0)\gamma^{\nu}\gR u(0)\,,
\end{equation}
in QCD and in the chiral limit, depends only on one invariant amplitude $\Pi_{LR}(Q^2)$ of the euclidean momentum 
squared $Q^2=-q^2$, with $q$ the momentum flowing through the two--point function:   
\begin{equation}\label{twopf}
\nonumber
2i\int d^4x e^{iq\cdot x} \langle 0\vert \mbox{T}\left\{L^{\mu}(x)\,R^{\nu}(0)
\right\}
\vert 0\rangle
\end{equation}
\vspace{-0.5cm}

\begin{equation}
\label{decomp}
\hspace{2cm}=\left(q^{\mu}q^{\nu}-q^2
g^{\mu\nu}\right)\Pi_{LR}(Q^2)\;.
\end{equation}

Here, we shall be particularly concerned with the study of $\Pi_{LR}(Q^2)$ in the limit of a large number of colors 
$N_c$ in QCD. We want to compute the lowest dimension operator product expansion (OPE) condensates of this 
function in two successive approximations to the large--$N_c$ limit: the so{-}called {\it minimal hadronic 
approximation} (MHA)~\cite{PPdeR01} consisting of a spectrum of the pion state, a vector state and an axial vector 
state; and the improved approximation where an extra higher  vector state is added.  

Part of our motivation comes from the fact that in the literature the values of these condensates are rather controversial.
Phenomenological analyses in refs.~\cite{PPdeR01,DGHS98,IZ01,Z04} find for example opposite sign for the first two 
condensates, in contradistinction to the results in refs.~\cite{BGP01,CGM03,RL04} which find the same sign. 

In this talk, we give results only for the two lowest dimension condensates and the matrix elements of $Q_7$ and 
$Q_8$ (see however \cite{FGdeR04} for more results and details).
  
In practice, we shall be working with the dimensionless complex function  $W_{LR}[z]$ defined as
\vspace{-0.5cm}
\begin{equation}\label{LR}
\nonumber
\hspace{-0.6cm}W_{LR}[{z} ]=-z\Pi_{LR}(z M_{\rho}^2)\,,\quad\with\quad \text{Re}\,z = \frac{Q^2}{M_{\rho}^2}\,,
\end{equation}
and use the mass of the lowest massive state, the $\rho(770~\MeV)$, to normalize quantities with dimensions. In 
large--$N_c$ QCD the function $ W_{LR}[z]$ is a meromorphic function and, therefore, in full generality, it can be 
approximated  by  successive  partial fractions of the type 
\begin{equation}\label{LRz}
W_{LR}[z]= A_{N}\prod_{i=1}^{P}\frac{1}{(z+\rho_{i})} \prod_{j=1}^{N}(z+\sigma_{j})\,, 
\end{equation}
with $\rho_1=1$ and $\rho_i \neq \rho_k$ for $i \neq k$;
where $P$ (and $N$) get larger and larger, but finite. $A_N$ is the overall normalization and
in what follows, we shall often use the notation $\rho_2 \equiv \rho_A$ and $\rho_3 \equiv \rho_{V'}$.

On the other hand, in QCD, the OPE of the two currents in Eq.~(\ref{decomp}) fixes the large--$Q^2$ fall off  in 
$1/Q^2$--powers of the invariant function $\Pi_{LR}(Q^2)$~\cite{SVZ79} to
\begin{equation}
\nonumber
\hspace{-0.6cm}
\Pi_{LR}(Q^2)\underset{Q^2\ra\infty}=\sum_{n=1}^{\infty}c_{2n+4}(Q^2,\mu^2)
\end{equation}
\vspace{-0.3cm}
\begin{equation}
\label{OPE}
\times\langle O_{2n+4}(\mu^2)\rangle(Q^2)^{-(n+2)}
=\frac{1}{2}
\sum_{n=1}^{\infty}
\frac{\langle \cO_{2n+4}\rangle}{(Q^2)^{n+2}} \,.
\end{equation}
Matching the leading asymptotic behaviour for large--$z$ in Eq.~(\ref{LRz}) to the one of the OPE in Eq.~(\ref{OPE}), 
restricts the number of zeros $N$ and the number of poles $P$ in Eq.~(\ref{LRz}) to obey the constraint
\begin{equation}
 N-P=-2\,.
\end{equation}
The case where $N=0$ corresponds to the MHA while the MHA+V' case is obtained for $N=1$.


\section{$\Pi_{LR}$ in the MHA and $\pi-V-A-V'$ Spectra}

It is claimed by some of the authors of refs.~\cite{BGP01,CGM03,RL04} that the reason why  their phenomenological
analysis of the chiral condensates  give the same sign for $\langle\cO_6\rangle$ and
$\langle\cO_8\rangle$ is due to the fact that the hadronic $\tau$--decay
spectrum is sensitive to the presence of the $\rho'$, while the MHA
ignores all higher states beyond the first axial state. Partly motivated by this claim, we want to analyze here the case, 
beyond the MHA, where an extra vector state $V'$, and therefore one zero $\sigma$, are also included. Let us collect 
some relevant equations
describing this case (the corresponding expressions in the MHA case can be obtained by putting $\rho_{V'}=\sigma$ 
in the MHA+V' equations).

\subsection{The Correlation Function}

With a spectrum of the pion pole, and $V$, $A$, and $V'$ states, the relevant correlation function is
\begin{equation}
\nonumber
W_{LR}[z]=A_1
\frac{z+\sigma}{(z+1)(z+\rho_A)(z+\rho_{V'})}\,,
\end{equation}
where \ \ $A_1\frac{\sigma}{\rho_A \rho_{V'}}=\frac{F_0^2}{M_V^2}\equiv \rho_F$\ \ \  and therefore
\begin{equation}
\label{condO6}
\langle\cO_6\rangle=\frac{-2}{\sigma}F_0^2 M_A^2 M_{V'}^2=-M_V^6 \frac{2}{\sigma} \rho_F \rho_A \rho_{V'}\,.
\end{equation}
Because of the positivity of $W_{LR} [z]$ for $\Ree\,z\ge 0$, the position of the zero
has to be in the Minkowski axis and, then, $\sigma>0$. 

\subsection{The Linear Constraint}

This is a very interesting constraint,  key equation of our analysis, which already provides a semi-quantitative 
argument in favor of the opposite sign option for the condensates $\langle\cO_{ 6} \rangle$ and $\langle\cO_8\rangle$. 
It simply
follows by expanding Eq.~(\ref{LRz}) to first non--trivial order in
inverse powers of
$z$
\begin{equation}\label{linear}
\sum_{j=1}^{N}\sigma_{j}-\sum_{i=1}^{P}\rho_{i}=\frac{1}{M_V^2}\frac{\langle
\cO_{8}\rangle}{\langle\cO_{6}\rangle}\,.
\end{equation}
 In the case corresponding to
the MHA, where by definition there are no zeros, this constraint
simply becomes:
\begin{equation}\label{O8O6}
1+\rho_{A}=-\frac{1}{M_V^2}\frac{\langle
\cO_{8}\rangle}{\langle\cO_{6}\rangle}\,,
\end{equation}
implying that, in the MHA  $\langle
\cO_{8}\rangle$ and $\langle
\cO_{6}\rangle$ must have {\it opposite signs}. 

Now, in the MHA+V' case,
Eq.~(\ref{linear}) reduces to 
\begin{equation}
\label{condO8}
\sigma-(1+\rho_A+\rho_{V'})=\frac{1}{M_V^2}\frac{\langle\cO_8\rangle}
{\langle\cO_6\rangle}\,.
\end{equation}
This equation  states 
that for
$\langle\cO_8\rangle$ to have the same sign as $\langle\cO_6\rangle$,
the position of the zero has to be {\it far beyond the largest
$V'$--pole}:
\begin{equation}\label{zp}
\sigma > 1+\rho_A+\rho_{V'}\,.
\end{equation}
Fixing the position of the poles at the values of the observed spectrum
(and ignoring errors for the purpose of the discussion){,} one has
$M_V={0.776}~\GeV$,
$M_A=1.230~\GeV \ (\rho_{A}=2.5)$ and
$M_{V'}=1.465~\GeV\ (\rho_{V'}=3.6)$;
which means that for the equal sign requirement option to be satisfied, one must have $\sigma>7.2$. In
$\GeV$ units this corresponds to a mass of $2.1~\GeV$. Now, in writing a large--$N_c$ ansatz for the $W_{LR}[z]$ 
function, one is implicitly assuming an effective cancellation between the extra poles and zeros in the complex 
$z$--plane which lie beyond a disc of radius $s_0$ covering all the poles and zeros retained in that approximation. 
The result $\sigma>7.2$
implies that the radius in question has to be $\sqrt{s_{0}}>2.1~\GeV$. A
priori that  {\it seems a good thing} because the OPE--matching is now applied at $Q^2\ge s_0$; i.e. in a
more asymptotic region than in the case of the MHA ansatz; however, it
also implies that there are no further poles in the region between  
$M_{V'}\simeq 1.5$ and the effective mass $M_{\sigma}\simeq 2.1~\GeV$ corresponding to the zero at $\sigma\simeq 
7.2$. This, however,  is in contradiction
with the observed $a_1$--like state at $M_{A'}\simeq 1.64 ~\GeV$ and $\rho$--like states at $M_{V''}\simeq 
1.72~\GeV$ and $M_{V'''}\simeq 1.9~\GeV$ below $M_{\sigma}\simeq 2.1~\GeV$. Alternatively, if one excludes those 
three states  $A'$, $V''$ and $V'''$ as all the phenomenological analyses using $\tau$--data do in fact, then the 
position of the zero $\sigma$ should be $\sigma\lesssim \frac{M_{A'}^2}{M_V^2}\approx4.5$, implying according to 
Eq.~(\ref{zp}), that $\langle \cO_8\rangle$ and $\langle \cO_6\rangle$ must have opposite signs, in contradiction with the 
claims of refs.~\cite{BGP01,CGM03,RL04}.

\section{Numerical Analyses and Conclusions}

Confronting the MHA and MHA+V' approximations to the experimental values of some observables that we introduce 
in the following will allow us to test their consistency and adjust their free parameters ($\rho_F$ and $\rho_A$ in the 
MHA case, $\rho_F$, $\rho_A$, $\rho_V'$ and $\sigma$ in the MHA+V' case) in order to make predictions for the 
condensates and matrix elements.
We use as input the following set of experimental data 
\begin{align}
&\hspace{-0.8cm}\delta m_\pi = 4.5936 \pm 0.0005 \;\MeV\,, &{\mbox{\rm ref.}}~\cite{PDG03}\nonumber\\
&\hspace{-0.8cm}L_{10} = (-5.13 \pm 0.19)\times10^{-3}\,, &{\mbox{\rm ref.}}~\cite{DGHS98} \nonumber\\
\nonumber&\hspace{-0.8cm}\Gamma_{\rho \ra e^+e^-} = (6.77 \pm 0.32)\times10^{-3} \;\MeV\,, &{\mbox{\rm 
ref.}}~\cite{PDG03}\nonumber
\end{align}
\begin{align} 
&\hspace{-0.8cm}\Gamma_{a \ra \pi\gamma} = (640 \pm 246)\times10^{-3} \;\MeV\,, &{\mbox{\rm 
ref.}}~\cite{PDG03}\nonumber\\
&\hspace{-0.8cm}\label{L9}L_9 = (6.9 \pm 0.7) \times 10^{-3}\,,&{\mbox{\rm ref.}}~\cite{BEG94}\nonumber\\
&\hspace{-0.8cm}M_{\rho}= (775.9 \pm 0.5)\;\MeV\,; &{\mbox{\rm ref.}}~\cite{PDG03}\nonumber
\end{align}

In fact, some of these observables, when expressed in terms of the MHA or MHA+V' parameters, depend not only on 
these ones but also on $(m_{\pi^+} + m_{\pi^0})$  and/or $M_V$. Therefore, it is more appropriate for our purposes to 
use in our fit procedure the dimensionless quantities: 
\begin{align}
&\frac{m_{\pi^+} + m_{\pi^0}}{M_\rho^2}\delta m_\pi = \frac{3}{4}\frac{\alpha}{\pi} \frac{\rho_A \log 
(\rho_A)}{\rho_A-1}\,,\\
&L_{10} =- \frac{1}{4} \rho_F \left( 1 + \frac{1}{\rho_A}\right)\,,\\
&\frac{1}{M_\rho}\Gamma_{\rho \ra e^+e^-} = \frac{4\pi\alpha^2}{3}\frac{\rho_A}{\rho_A-1}\,,\\
&\frac{1}{M_\rho}\Gamma_{a \ra \pi\gamma} = \frac{\alpha}{24} \frac{\sqrt{\rho_A}}{\rho_A-1}\,,\\
&\label{L9}L_9 = \frac{1}{2}\rho_F\;.
\end{align}
For lack of place we gave the formulae only in the case of the MHA, the corresponding MHA+V' equations can be found 
in \cite{FGdeR04}.

A standard $\chi^2$ statistical regression method permits to fit the parameters.

\subsection{The case of the MHA spectrum}
For this spectrum, we find $\rho_F=(12.36 \pm 0.35)\times 10^{-3}$\ \  and \ \ $\rho_A=1.464 \pm 0.004$,        
with a $\chi^2_{\text{min}}=1.21$ for 3 degrees of freedom (dof). The covariance matrix is given by
\begin{equation}
\text{cov} \left(\rho_F, \rho_A\right) = 
\begin{pmatrix}
1.21 & 1.36 \\
1.36 & 162 
\end{pmatrix}\times 10^{-7}\;.
\end{equation}

We deduce from these results two conclusions: first that the MHA framework is statistically relevant and second that 
the fitted free parameters have small statistical errors.

\subsection{The case of a $\pi - V - A - V'$ spectrum}

Eq.~(\ref{L9}) becomes now a function of $g_V$ as explained in 
\cite{FGdeR04}. The way we treat this is by considering the observable\footnote{The decay width of $\rho \rightarrow 
\pi\pi$ is $\Gamma_{\rho \rightarrow \pi\pi}=(150.4\pm 1.3) \MeV$ \cite{PDG03}.}:
\begin{equation}
\nonumber
\hspace{-0.8cm}\frac{1}{M_\rho}\Gamma_{\rho\rightarrow \pi\pi} = \frac{1}{48\pi} 
\end{equation}

\vspace{-0.7cm}

\begin{equation}
\nonumber
\hspace{0.3cm}\times\frac{1}{\rho_F^2}\frac{\sigma \left(\rho_F - 2 L_9 
\rho_{V'}\right)^2(\rho_A-1)(\rho_{V'}-1)}{(1-\rho_{V'})^2\rho_F\rho_A \rho_{V'}(\sigma-1)}\;,
\nonumber
\end{equation}
as a function of $L_9$ which has an error itself, and is added as an extra parameter in our fit. The number of dof does 
not change since $L_9$ is also taken as an observable.
For this case, we impose a criterion of rejection through the 
ordering: 
\vspace{-0.2cm}
\begin{equation}
\nonumber
\rho_A<\rho_{V'}<\sigma < \rho_0 \doteq \frac{s_0}{M_V^2} \;.   
\end{equation} 

\vspace*{-0.2cm}

The first and second inequalities reflect the knowledge that the new state has a higher mass than the axial, that it is a 
$V$--like pole and, therefore, its residue contributes positively to the $W_{LR}(z)$--function; the third inequality 
follows from the requirement  that the perturbative threshold $s_0$ already lies  beyond the 
radius where the analytic structure of the poles and zeros retained satisfies the leading OPE constraint.

The $\chi^2$ regression leads to 
{\setl
\bea
\rho_F & = & (12.36 \pm 0.03)\times 10^{-3}\,, \label{rhoF} \\
\rho_A & = & 1.466 \pm 0.003\,, \\
\rho_{V'} & =  & 2.63 \pm 0.01\,, \\
\sigma & = & 2.64 \pm 0.01\,. \label{sigma}
\eea}
The results in Eqs.~(\ref{rhoF}) to (\ref{sigma}) correspond to a value:  $L_9=(6.44 \pm 0.02)\times 10^{-3}$, with a 
$\chi^2_{\text{min}}=0.60$ for $1$ dof. 

We conclude that the parameters $\rho_F$ and $\rho_A$ are statistically stable when compared to those found in the 
MHA case.
We also find that $\rho_{V'} \approx \sigma$, which is consistent with the fact that the MHA approximation already seems to 
have the bulk of the full large--$N_c$ information. In other words, adding an extra  V'--pole appears to be 
compensated, at a very good approximation,  by the position of the nearby zero. 

\subsection{predictions}
Using Eqs. (\ref{condO6}), (\ref{condO8}) (and corresponding equations in the MHA case) as well as expressions 
of the matrix elements of four-quark operators (\cite{FGdeR04},\cite{KdeR97}), we find

\vspace{0.3cm}

\hspace{-0.9cm}
\begin{tabular} [c] {|c|c|c|}
\hline
 & $\left< \mathcal{O}_6\right>$ & $\left< \mathcal{O}_8\right>$  \\
 & $\times 10^3 \; \; \text{GeV}^6$ & $\times 10^3\; \; \text{GeV}^8$\\
\hline
MHA + V' & $-7.90 \pm 0.20 \pm1.62$ & $ +11.69 \pm 0.32 \pm2.53$ \\
\hline 
MHA & $-7.89 \pm 0.23 \pm2.01$ & $+11.71 \pm 0.34 \pm3.08$ \\
\hline
\end{tabular}

\vspace{0.3cm}

\hspace{-0.9cm}
\begin{tabular} [c] {|c|c|c|}
\hline
 & $M_7$ (NDR) &  $M_7$ (HV)  \\
\hline
MHA + V' & $0.12 \pm 0.00 \pm 0.01 $ & $0.59 \pm0.01 \pm0.06$  \\
\hline 
MHA & $0.12 \pm 0.00 \pm 0.02$ & $0.59 \pm 0.01 \pm 0.11$  \\
\hline
\end{tabular}

\hspace{-0.9cm}
\begin{tabular} [c] {|c|c|c|}
\hline
 & $M_8$ (NDR) &  $M_8$ (HV)  \\
\hline
MHA + V' &  $2.00 \pm 0.03 \pm 0.20$ & $2.15 \pm 0.03 \pm0.22$ \\
\hline 
MHA & $1.99\pm 0.03 \pm 0.36$ & $2.15 \pm 0.03 \pm 0.39$ \\
\hline
\end{tabular}

\vspace{0.3cm}

The way of obtention of these results as well as the origin of systematic errors are explained in \cite{FGdeR04}.

We then see that within errors, the two sets of predictions from MHA and from MHA + V' are perfectly consistent with 
each other.

\end{document}